\documentclass[twoside]{article}

\usepackage{graphicx}
\usepackage{epic}
\usepackage{amsmath}
\usepackage{amssymb}
\usepackage{macrosv1}
\usepackage{macros_static}
\usepackage{macros_add}
\usepackage{epsfig}
\usepackage{fleqn,espcrc2}
\usepackage[dvips]{feynmp}

\newcommand{\ftext}[1]{\fbox{ {#1} }}

\title{
{
\vspace{-3.0cm} \normalsize \hfill
\parbox{30mm}{ DESY 02-132 \\
               September 2002 
             } 
}\\[20mm]
Non-perturbative renormalization of HQET and 
QCD\thanks{Plenary talk at ``Lattice 2002'', Boston MA.}
}

\author{Rainer Sommer\address{DESY, Platanenallee 6, 15738 Zeuthen, Germany}}

\begin{document}

\begin{abstract}
We discuss the necessity of non-perturbative renormalization
in QCD and HQET and explain
the general strategy for solving this problem. 
A few selected topics are discussed in some detail, namely the importance of 
off-shell improvement in the MOM-scheme on the lattice, recent progress
in the implementation of finite volume schemes and then 
particular emphasis is put on the recent
idea to carry out a non-perturbative renormalization of the
{\em Heavy Quark Effective Theory} (HQET).
\end{abstract}

\maketitle
\section{INTRODUCTION \label{s_intro}}

Non-perturbative renormalization is an important mile-stone
on the path from lattice QCD computations to precise
predictions for particle physics phenomenology. Why is this so?
One might think that only
the QCD coupling  and quark masses need to be renormalized.
However, very important phenomenological 
applications of lattice QCD concern the physics of heavy quarks
and  weak decays.
As illustrated in \fig{f_scales}, these involve energy scales (masses), 
which are far too high
for the lattices that can be simulated on present and (near future) 
computers. Effective theories have to be used,
summarizing the effects of the heavy fields, which can't
be treated dynamically, in local 
composite operators. They are a way to
implement expansions in terms of $1/M_x,\, x={\rm W},{\rm t},\ldots$.
The requirement that the corrections to the $n$-th order 
are really $(1/M_x)^{n+1}$ and terms such as $\alpha_{\rm s}/(M_x)^n$ do not 
appear fixes the renormalization of the operators
in the effective Lagrangian. We denote the so-defined 
renormalization scheme by ``match''.

To give a specific example, the 
mixing amplitude between a K-meson and a $\overline{\rm K}$
is to lowest order in the weak interaction given
by a QCD matrix element of a 4-fermion operator whose
(re)normalization is completely fixed by the matching
to the standard model amplitude. 

Of course, already before considering weak decays,
the renormalization of the QCD coupling
and quark masses is important to obtain the basic
renormalization group invariants (RGI) of the theory:
the Lambda parameter, $\Lambda_\msbar$, and the RGI quark 
masses, $M_i,i={\rm u},{\rm d},\ldots$. Their definition
and their special r\^ole in parameterizing QCD is briefly 
but thoroughly explained in
\cite{lat02:francesco}.
 
\begin{figure}[ht]
    \vspace{0 cm}
    \hspace{0 cm}
    \includegraphics[width=6cm]{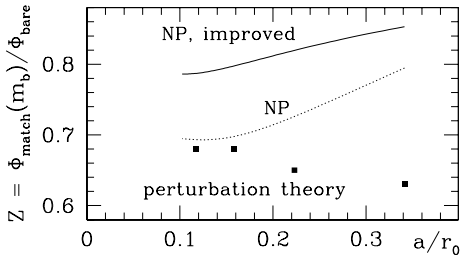}
    \vspace{-0.5 cm}
\caption{\small \label{f_zastat}
Non-perturbative (NP) 
renormalization constant of the 
static-light axial current~\cite{lat02:jochen} and
its estimate\cite{stat:fnal2} from (tadpole-improved~\cite{Lepenzie}) 
perturbation theory. The label ``improved'' refers to the 
$\Oa$-improved theory.
}
\end{figure}

Why should we do renormalization non-perturbatively? There are
good reasons:\\[0.5ex]
1. Concerning $\Lambda_\msbar, M_i$, 
phenomenological determinations claim very good precision 
but are frequently based on difficult to test assumptions,
most notably the applicability of perturbation theory
in various kinematical regions. 
In my opinion, the r\^ole of lattice QCD is complementary: 
to use as few assumptions as possible. 
This means to
perform renormalization {\em non}-perturbatively.\\[0.5ex]
2. Some examples demonstrate that perturbative
renormalization is not sufficiently precise in practice.
E.g. the JLQCD collaboration computed $B_{\rm K}$ with two 
different perturbatively renormalized operators which should
be equivalent in the continuum limit.
Performing the continuum extrapolations, they
could only find agreement 
by fitting $\alpha^2$ corrections in addition to
the $a^2$ lattice artifacts~\cite{bk:JLQCD}. A more recent 
example is the static-light axial current.
As shown in \fig{f_zastat} perturbative and non-perturbative 
Z-factors differ by more than the unavoidable $\Oa$ term.



\newcommand{\cred}{}
\newcommand{\cblu}{}
\newcommand{\cmag}{}
\newcommand{\cgre}{}
\newcommand{\cbla}{}

\newcommand{\mgt}{\cmag}

\newcommand{\logenergy}[1] 
{
\unitlength #1
\linethickness{0.3mm}
\mgt\put(1.0,0.0){\line(1,0){18}}
\put(17,-1.0){ Energy, $\mu$ }
\mgt{\dottedline{0.2}(18,0)(21,0)}
\put(21,-1.0){\cbla $\infty$}

\multiput(2,0)(4.0,0){4}{\line( 0, -1){0.5}}
\put(21,0){\line( 0, -1){0.5}}
\cbla 
\put(1,-1){\small 100 MeV}
\put(5.2,-1){\small 1 GeV}
\put(9.1,-1){\small 10 GeV}
\put(13.2,-1){\small 100 GeV}
}
\newcommand{\logmu}[1] 
{
\unitlength #1
\linethickness{0.3mm}
\mgt\put(1.0,0.0){\line(1,0){18}}
\put(17,-1.0){\large $\mu$}
\mgt{\dottedline{0.2}(18,0)(21,0)}
\put(21,-1.0){\cbla $\infty$}

\multiput(2,0)(4.0,0){4}{\line( 0, -1){0.5}}
\put(21,0){\line( 0, -1){0.5}}
\cbla 
\put(1,-1){\small 100 MeV}
\put(5.2,-1){\small 1 GeV}
\put(9.1,-1){\small 10 GeV}
\put(13.2,-1){\small 100 GeV}
}

\newcommand{\linenergy}[1] 
{
\unitlength #1
\linethickness{0.3mm}
\mgt\put(1.0,0.0){\line(1,0){17}}
\put(14,-1.0){ Energy }
\put(15,-2.0){ [GeV] }

\multiput(1,0)(2.0,0){7}{\line( 0, -1){0.5}}
\cbla
\put(0.9,-1.2){\small 0}
\put(4.9,-1.2){\small 4}
\put(8.9,-1.2){\small 8}
\put(12.5,-1.2){\small 12}
}

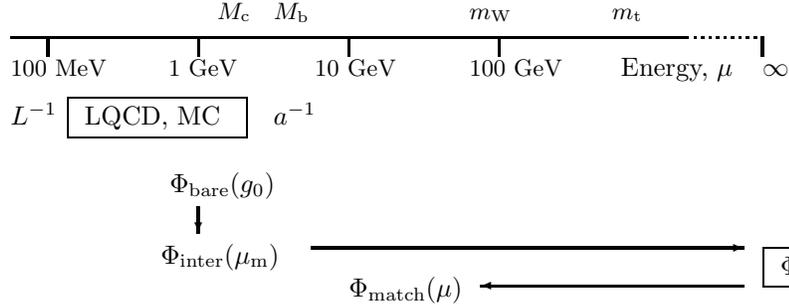
\begin{figure*}
\hspace{1.5cm}
\begin{picture}(18,9)(0,30)
  \unitlength 0.5cm
  \put(0,2){\logenergy{0.5cm}}
  \put(6.5,2.5){\small $\Mc$} 
  \put(8,2.5){\small $\Mb$}
  \put(13.2,2.5){\small $m_{\rm W}$}
  \put(17,2.5){\small $\mt$}
  \put(1,-0.3){\cred $L^{-1}$}
  \put(2.5,-0.3){\ftext{LQCD, MC }}
  \put(8,-0.3){\cred $a^{-1}$} 
  \cbla\put(5,-2.1){ 
                  $ \Phi_\bare(g_0)$
                }
\linethickness{0.4mm}
  \cgre\put(6,-2.5){\vector(0,-1){0.7}}
  \cred \put(5,-4.0){$\Phiinter(\mu_{\rm m})$}
\linethickness{0.3mm}
  \put(9,-3.6){\vector(1,0){11.5}}
  \cmag \put(21,-4.3){\ftext{$\PhiRGI$}}
\linethickness{0.3mm}
  \cgre \put(10,-4.9){$\Phimatch(\mu)$}
\linethickness{0.3mm}
  \put(20.5,-4.6){\vector(-1,0){7.0}}
\end{picture}
\vspace{2.9cm}
\caption{\small \label{f_scales}
Top: The standard model summarizes physics at many different scales,
while lattice QCD simulated by Monte Carlo (MC) can correctly cover only
a small range between the inverse lattice spacing $a^{-1}$, and the 
box size $L$. 
Bottom: The strategy to connect the bare matrix elements to the ones in the 
matching scheme.
}
\end{figure*}
 
\section{GENERAL STRATEGY \label{s_strat}}
We now discuss the implementation of scale-dependent
renormalizations, leaving aside simpler cases 
such as the renormalization of the axial 
current~\cite{za:comb}, where the (scale independent)
renormalization may be determined from a symmetry. Focusing on multiplicative 
renormalization, the bare operators,
$\op{\bare}$, and the renormalized ones
at scale $\mu$, $\op{\inter}(\mu)$, are related by
\bes
  \label{e_Zinter1}
  \op{\inter}(\mu) = \zinter(g_0,a\mu) \op{\rm bare}\,, 
\ees
with a renormalization factor $Z$ (or a matrix) depending
on the bare coupling, $g_0$, and the combination $a\mu$. In 
an {\em intermediate renormalization scheme}, $Z$ is fixed by requiring
\bes
  \label{e_Zinter2}
   \langle \beta| \op{\inter}(\mu) |\alpha \rangle = 
   \langle \beta| {\cblu \op{\bare}} |\alpha \rangle_{\rm \cblu tree level} 
\ees
for convenient states $|\alpha \rangle, |\beta \rangle$. 
To have simple renormalization group equations (RGE), it is important
that $|\alpha \rangle, |\beta \rangle$ are characterized through
one scale, $\mu$, only. In the two most frequently
used intermediate schemes, this is realized as follows. \\ 
SF: In the Schr\"odinger functional 
schemes~\cite{impr:lett}, \eq{e_Zinter2} is formulated 
through the gauge invariant path integral in a finite space-time
volume $T\times L\times L\times L$ with $T/L$
fixed. $|\alpha \rangle, |\beta \rangle$ are given in terms of 
some boundary states (propagated in Euclidean time) 
and one has $\mu=1/L$.\\
MOM~\cite{RIMOM}: Here
one uses infinite volume quark states in Landau
gauge, with four-momenta $p$ of the quarks satisfying $p^2=\mu^2$.

Both are massless schemes, their renormalization
constants are evaluated at zero quark masses. 
In addition, the renormalized operators are
independent of the regularization.\footnote{Thus also 
the name RI/MOM is used instead of MOM.} More precisely, 
(the limit $a\to0$ of) the matrix elements
\bes
 \Phiinter(\mu) \equiv \langle f | \op{\inter}(\mu) | i \rangle
\ees
are independent of the regularization and the perturbative
coefficients $\gamma_i$ in the RGE
\bes
  \label{e_RG}
  \mu {\rmd \Phiinter(\mu) \over \rmd \mu}  &=&\gamma(\gbar^2(\mu)) \,
  \Phiinter(\mu) \\
  \gamma(\gbar^2) &=& - [\gamma_0 \gbar^2 + \gamma_1 \gbar^4 + \ldots]
  \label{e_RG_pert}
\ees
can in principle be computed using dimensional or
a lattice regularization.

\begin{figure*}[t]
\begin{center}
    \vspace{0.5 cm}
    \hspace{0 cm}
    \includegraphics[width=11cm]{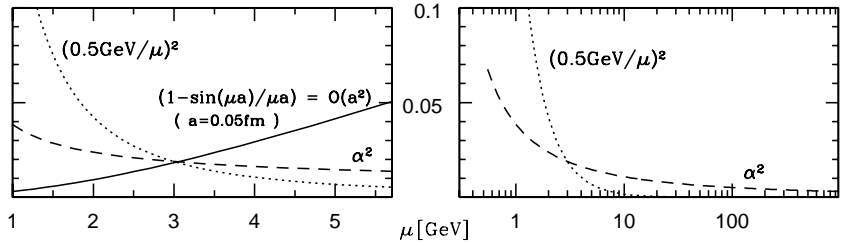}
    \vspace{-1.2 cm}
\end{center}
\caption{\label{f_challenges}
\small Typical relative error terms 
as they show up 
in infinite volume schemes (left) 
and finite volume schemes (right): $a^2$-effects as well as
non-perturbative (dotted) and perturbative (dashed) terms. 
 }
\end{figure*}
For phenomenological applications we are interested 
in the matrix elements $\Phimatch(\mu)$ in the 
matching scheme mentioned in \sect{s_intro}. To connect to it,
it is usually\footnote{For an alternative see \sect{s_HQET}.}
 convenient to first compute the {\em scheme independent}
renormalization group invariant (RGI)
\bes 
  \label{e_RGI}
  \PhiRGI  
    &=& \lim_{\mu \to \infty} 
        \Phiinter(\mu) \left[2 b_0\gbar^2(\mu)\right]^{-\gamma_0/2b_0^2} \\
  &=& \Phiinter(\mu) (2 b_0\gbar(\mu)^2)^{-\gamma_0/2b_0^2} \label{e_RGI2} \\
          &&\times  \exp \left\{- \int_0^{\gbar(\mu)} \rmd g [\frac{\gamma(g)}{\beta(g)}
            - \frac{\gamma_0}{b_0 g} ] \right\} \nonumber
\ees
and then connect $\PhiRGI$ to $\Phimatch(\mu)$ 
through the analogue of eqs.(\ref{e_RG},\ref{e_RGI}) for 
the matching scheme. This last step is mostly 
done by perturbation theory and -- as discussed by P. Hasenfratz at last year's
conference~\cite{lat01:hase} -- its reliability needs to be 
investigated. For the purpose of this report, we note that the RGI
matrix elements~\eq{e_RGI} are the fundamental quantities of QCD and should 
be determined from lattice QCD with good precision. 

The strategy is sketched in the bottom part of \fig{f_scales}.
For multiplicative renormalization, each arrow in this graph is 
realized by a factor,
\bes
   \label{e_factors}
   \Phimatch(\mu) &=& {\Phimatch(\mu) \over \PhiRGI} 
                    \times
                    {\PhiRGI \over \Phiinter(\mu_{\rm m})} \times  \\ 
                  &&  \zinter(g_0,\mu_{\rm m} a) \times \Phibare(g_0)\,. \nonumber
\ees
The first two factors are independent of the lattice regularization
(e.g. choice of action).
\subsection{\bf Challenges} 
The practical implementation of this programme represents a challenge 
beyond the one present in 
spectrum calculations, since the limit $\mu \to \infty$
in \eq{e_RGI} has to be controlled. 

With MOM
(or other infinite volume schemes) as an intermediate scheme, one
computes $\Phiinter(\mu)$  with $\zinter(g_0,a\mu)$ from 
eqs.(\ref{e_Zinter1},\ref{e_Zinter2}) for $\mu=\rmO(2\,\GeV)$
and uses the {\em perturbative} RG \eq{e_RG_pert} inserted
into \eq{e_RGI2} to continue to 
$\mu \to \infty$. Systematic errors of the
order shown on the l.h.s. of \fig{f_challenges} have to
be expected. It is evidently not easy to disentangle
the various error sources in the window
of $\mu$ and $a$ available on large volume lattices.

To cleanly separate the lattice artifacts ($\rmO(a^2)$ or $\Oa$)
from the physical $\mu$-dependence, 
renormalization conditions may also be posed in finite volume 
(e.g. SF-scheme) and then a recursion $\mu \to 2\mu$ allows
to reach large enough $\mu$
to make sure that the terms illustrated
on the r.h.s. of \fig{f_challenges} are 
negligible/controlled~\cite{alpha:sigma,mbar:pap1}. 
The key point is that
before comparing the $\mu$-dependence to
the perturbative one, 
the $a$-effects can be controlled
in each step of the recursion. Examples are given in \sect{s_finite_vol}.


\section{SELECTED TOPICS \label{select}}
We now discuss some selected topics of relevance. Recent reviews,
with partly different emphasis, are 
\cite{reviews:renorm}.

\subsection{MOM-scheme\label{s_MOM}}
The MOM-scheme~\cite{RIMOM} is very popular.
It is rather easy to
implement and does not require simulations to be done specifically
for the renormalization. A number of new numerical results
have been presented at this 
conference~\cite{lat02:reyes,lat02:lubicz,lat02:laurent}.
Here 
we  concentrate on two aspects which have not been
discussed much.
\subsubsection{$\Oa$-improvement \label{s_offshell}}
\begin{figure*}
  \begin{center}
  \epsfig{file=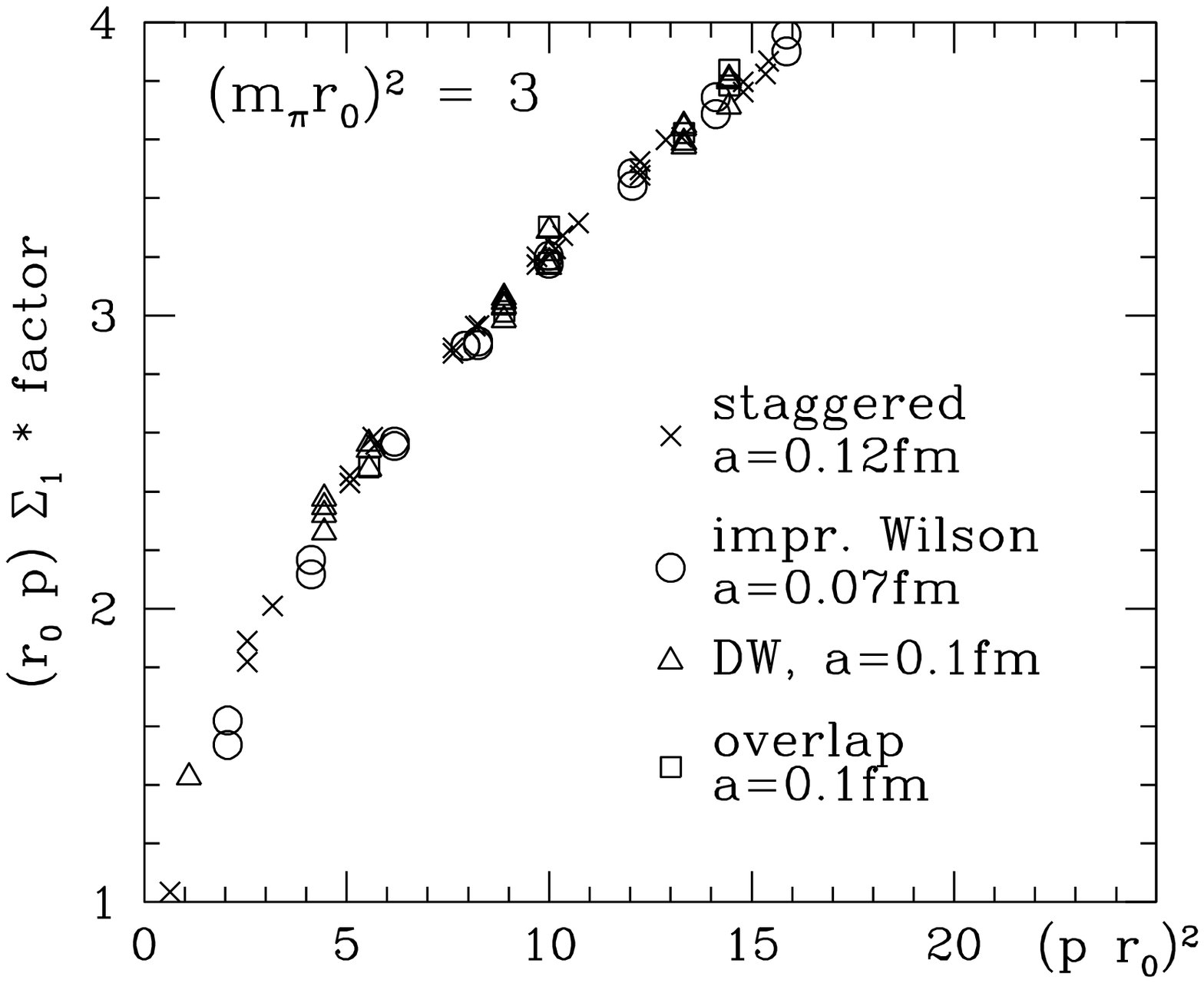,width=6.5cm} \hspace{0.8cm}
  \epsfig{file=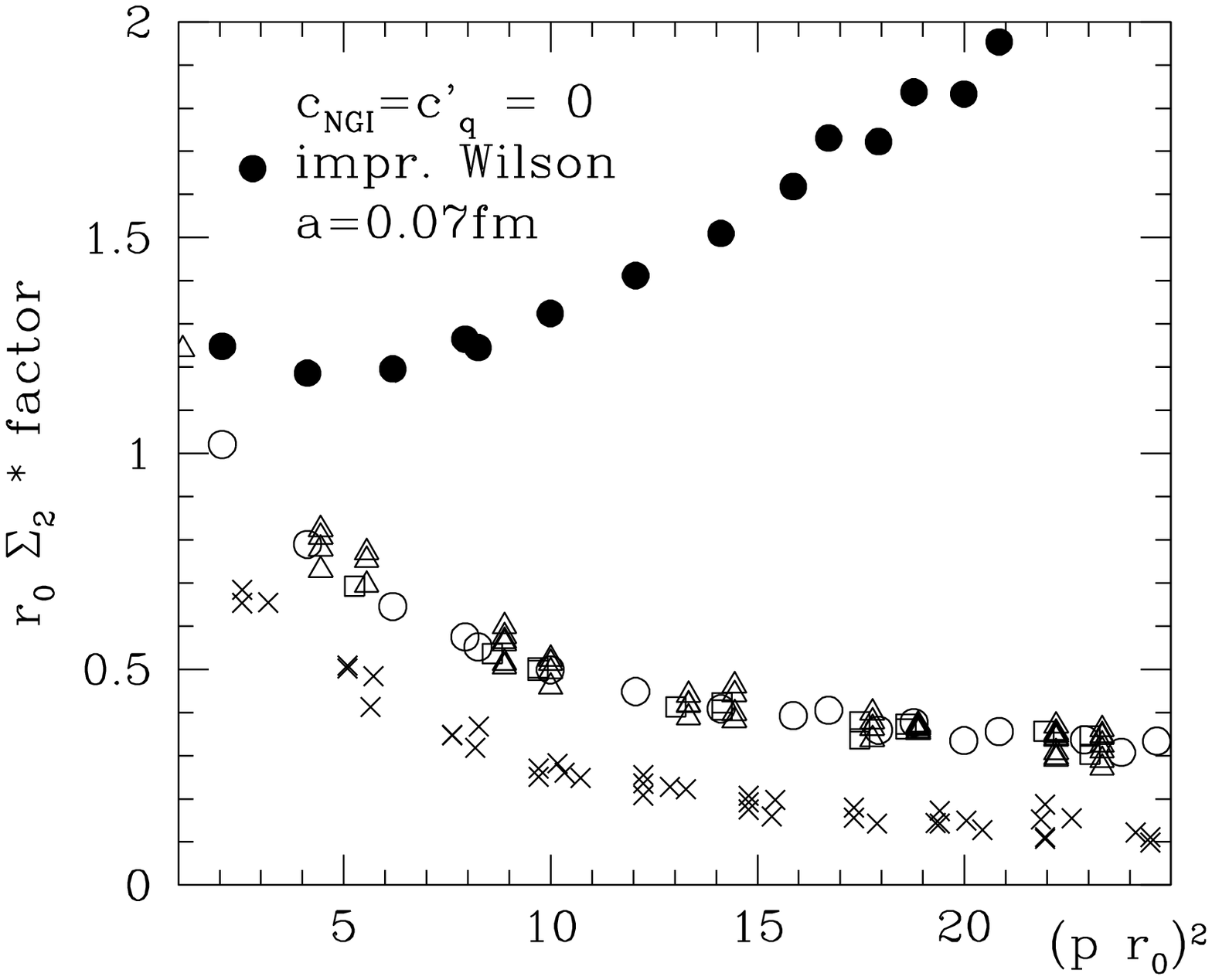,width=6.5cm}
    \vspace{-1 cm}
  \end{center}
\caption{\small \label{f_qprop}
 The functions $\Sigma_i$, \protect\eq{e_qprop2},
 renormalized such that they coincide at $\Sigma_1(r_0^2p^2=5)$. 
 Results are from \cite{qprop:GW} (overlap), \cite{qprop:DW} (DW), 
 \cite{lat01:bhatta} (improved Wilson) and \cite{qprop:KS} (staggered).
 Open circles use estimates of  $c'_{\rm q},c_{\rm \small NGI}$ \cite{lat01:bhatta},
 while filled ones  are for $c'_{\rm q}=c_{\rm \small NGI}=0$.
 Error bars are comparable to symbol sizes. 
}
\end{figure*}

Renormalization conditions in the 
MOM-scheme are based on off-shell quark 
Greens-functions in 
Landau gauge, which {\em are not} $\Oa$-improved by the standard
on-shell improvement programme~\cite{impr:comb,impr:pap3}. 
The most basic ingredient is the momentum space quark
propagator
\bes
  \label{e_qprop1}
  \tilde{S}(p) &=& \sum_x \rme^{-ipx} \langle \psi(x)\psibar(0)\rangle \\[-0.5ex] 
   &\equiv& [i \gamma_\mu p_{\mu} \Sigma_1 + \Sigma_2]^{-1}\,. 
  \label{e_qprop2}
\ees
For sufficiently large $p^2$, it is computable in perturbation
theory and $\Sigma_2$ 
is expected to go to a constant (asymptotic freedom!).
Evaluating $\Sigma_2$ with the non-perturbatively on-shell improved
action~\cite{impr:pap3} (e.g. at $a=0.07\,\fm$), 
one finds instead a strong rise 
with $p^2$~\cite{qprop:p1,lat01:bhatta}
($\bullet$ in \fig{f_qprop}). The SPQ$_{\rm CD}$R Collaboration
verified that this lattice artifact is reduced for
decreasing $a$~\cite{lat02:lubicz}. We learn that off-shell
$\Oa$ lattice artifacts are large and their impact
on the numerical results in the MOM-scheme may be substantial.

The possibilities of reducing them in a 
systematic Symanzik improvement programme were studied in
\cite{impr:offshell1,impr:offshell2}. A rough summary follows.

Since one is interested in non gauge invariant Greens functions,
the improvement terms have to satisfy only
BRST invariance instead of gauge invariance.
In addition,
contact terms have to be subtracted from the
correlation functions in position space. 

We restrict ourselves to the quark propagator and  
correlation functions of the form 
$\langle \psi(x) \psibar(y) \op{\Gamma}(z) \rangle$,
with $\op{\Gamma}=\psibar \Gamma \tau^a \psi$,
$\Gamma$ a matrix in Dirac-space and
a traceless matrix $\tau^a$ in flavor space. 
These are the Green's functions needed for the MOM-renormalization of
flavor non-singlet 2-fermion operators.
Then off-shell $\Oa$-improvement
may be achieved by~\cite{impr:offshell1}\\[1ex]
1) replacing the quark field $\psi(x)$ with 
\bes
 {\cmag \psi_{\rm R}(x)} &=& Z_{\rm q}^{-1/2}\, (1-\frac12 b_{\rm q} am_{\rm q}) \,
 \psi_{\rm I}(x)
 \\
 \psi_{\rm I}(x) &=& [1+ac'_{\rm q} D + a c_{\rm \small NGI}
                            \partial_\mu\gamma_\mu] \psi(x)\,, \nonumber
\ees
with $g_0$-dependent coefficients 
$Z_{\rm q},b_{\rm q},c'_{\rm q},c_{\rm \small NGI}$ (and similarly for
$\psibar$).\\[1ex] 
2) adding a term
$ac'_{\Gamma}\psibar \Gamma D \psi$ to $\op{\Gamma}$
in addition to the on-shell improvement terms of \cite{impr:comb}.

Above, $D$ denotes the full lattice Dirac operator
including the mass term. An important point to note is that this is strictly 
speaking {\em not an improvement of the fields}: 
 e.g. an $N$-point function of $\op{\Gamma}(x)$
will be $\Oa$-improved when considered on-shell (all operators
separated by a physical distance) but not when considered off-shell
(e.g. Fourier transformed). In the latter case, 
additional contact terms have to be subtracted in general.
Nevertheless, the
above is a possible form of writing the $\Oa$ counter-terms
needed for the MOM-scheme renormalization
of $\op{\Gamma}$~\cite{impr:offshell1}.

Investigations of practical ways to determine 
the new coefficients $c'_{\rm q},c_{\rm \small NGI}$ have started
\cite{lat01:bhatta,lat01:sharpe}. Before we discuss the result,
we turn to off-shell improvement in the context of
formulations of QCD with exact chiral 
symmetry (see \cite{lat02:giusti}). Their
massless Dirac operator $D$ satisfies the Ginsparg-Wilson relation~\cite{GW}
\bes
  \gamma_5 D^{-1} + D^{-1} \gamma_5 = a \,2R\,\gamma_5\,, \label{e_GW}
\ees 
with a local operator $R$ 
and the fermionic action
is $S=a^4\sum_x\psibar(x)[D+m_0(1-\frac{a}{2}RD)]\psi(x)$.\\ 
Examples of {\em local Dirac operators}
satisfying \eq{e_GW} are known \cite{exactchi:neub,exactchi:perfect}
and in addition it has been shown that
Domain wall (DW) fermions~\cite{exactchi:furshamir} are 
approximate realizations~\cite{exactchi:kikunogu}.

A possibly confusing point is that the MOM-scheme
is applied for these formulations, stating that there is 
$\Oa$-improvement because of chiral symmetry,
while L\"uscher's exact chiral symmetry~\cite{exactchi:martin} is
only valid on-shell (otherwise it would be in 
contradiction to the Nielsen-Ninomia theorem~\cite{NiNi}). 
We briefly explain in which sense off-shell $\Oa$-improvement
is implied by \eq{e_GW}; 
see \cite{impr:offshell3} for a similar discussion.
For simplicity we assume that $R$ is proportional to unity.

Introduce quark fields and anti-quark fields 
\bes
  \label{e_imprfields}
  \psi_{\rm I}(x) = (1 - \frac{a}{2}RD)\psi(x) \,, \quad
  \psibar_{\rm I}(x) =  \psibar(x)\,,
\ees
and perform the same replacement in quark bilinears,
\bes
  \op{\Gamma,\rm I}(x) = \psibar_{\rm I}(x) \Gamma \tau^a \psi_{\rm I}(x)\,. 
  \label{e_Pimpr}
\ees
It follows immediately that the quark propagator
\bes
  \tilde{S_{\rm I}}(p) &=& 
   \sum_x \rme^{-ipx} \langle \psi_{\rm I}(x)\psibar_{\rm I}(0)\rangle
\ees
satisfies $\tilde{S_{\rm I}}(p)\gamma_5 + \gamma_5 \tilde{S_{\rm I}}(p)=0$ 
for $m_0=0$;
the continuum relation is exact at finite lattice spacing. Furthermore,
the Green's functions of quark bilinears, \eq{e_Pimpr}, 
satisfy the continuum chiral Ward identities 
at finite lattice spacing. This excludes the necessity of
additional $\Oa$ improvement terms since all 
possible terms~\cite{impr:offshell1} transform differently under chiral rotations
and would lead to violations of the Ward identities.
One concludes that the off-shell improved quark fields and quark bilinears 
are known without any free improvement coefficients. 

Note that the action is {\em not} written in terms of the fields 
$\psi_{\rm I}$, 
$\psibar_{\rm I}$ since this would lead to a non-local 
Dirac operator (in agreement with the Nielsen-Ninomia theorem);
only the fields in the correlation functions are improved. 
Concerning DW fermions in the limit
of large ``fifth dimension'', the situation was clarified by 
Kikukawa and Noguchi. They derived a 4-d effective action 
with fields $\psi,\psibar$ and
a local Dirac operator which satisfies \eq{e_GW}
with $R=2$.
The Shamir-Furman fermion fields\cite{exactchi:furshamir}, 
which are constructed from the 5-d fields at the boundaries,  
are identified with $\psi_{\rm I}$, 
$\psibar_{\rm I}$ exactly related to $\psi,\psibar$ by \eq{e_imprfields}.

We now come back to numerical results. 
In \fig{f_qprop}, $\Sigma_1$ and $\Sigma_2$ are displayed for various 
discretizations. The results for DW fermions~\cite{qprop:DW} 
and overlap fermions~\cite{qprop:GW}
show the expected leveling off
of $\Sigma_2$ with $p^2$ and are in complete agreement with each other
within their (not so small) statistical errors. 
As discussed before, $\Oa$-errors are expected to be absent\footnote{
       We assume that the extent of the fifth dimension in the 
       DW computation was large enough.}. 
The mutual agreement is an indication that $\rmO(a^2)$-errors are small.
More surprisingly, also the approximately off-shell $\Oa$-improved 
Wilson fermion results~\cite{lat01:bhatta} fit in very well! 
On the other hand, the staggered fermion $\Sigma_2$~\cite{qprop:KS} 
appears to have significant discretization errors at $a=0.12$~fm.

What does this mean for NP renormalization in the MOM-scheme? 
Since the renormalization conditions, \eq{e_Zinter2}, are imposed 
in such a way that $Z=1+\rmO(g_0^2)$
holds exactly at finite $a$, discretization
errors are suppressed by one order of $g_0^2$. 
Whether this is sufficient needs more
investigations.
In any case, formulations with exact 
chiral symmetry
~\cite{qprop:DW,qprop:GW} are expected to be superior and
first numerical  results for MOM-scheme 
$Z$'s~\cite{qprop:DW,qprop:GW} appear to confirm this.

At this conference, new results for MOM renormalization constants
in the on-shell improved theory were 
presented~\cite{lat02:reyes,lat02:lubicz}. Not all of them are
well compatible with the perturbative RG to the NL order. It remains
to be clarified why this is so.
Apart from a true non-perturbative $\mu$-dependence,
$\Oa$ errors or Goldstone pole contributions
are possible explanations.

\subsubsection{Goldstone poles}
To discuss the latter possibility, we choose the simplest case, 
$\op{5}=\psibar\gamma_5\tau^a\psi$ and write schematically  
\bes
  R(m) = {\langle p' | \op{5} | p \rangle \;/\;  
  \langle p' | \op{5} | p \rangle_{\rm \cblu tree level}}\,,
\ees
with $\langle p' |$  ($| p \rangle$) an (anti-) quark state.
It is
standard to choose the forward kinematics $p=p'$
and $\zp^{-1}$ in the MOM-scheme is to be obtained from $R(m)$
in the limit where the quark mass $m$ goes to zero. 
However, it is well known (see e.g. \cite{RIMOM,goldstonepole}) that
$R(m)$ is singular in the chiral limit due to the 
coupling of the pion to the pseudo-scalar vertex and
the fact that $p-p'=0$ in the chosen kinematics; one expects  
\bes
  \label{e_Rm}
  R(m) = A(p^2) {1 \over m^2_\pi(m)} + B(p^2) + \rmO(m_\pi^2)\,.
\ees
Based on this form it has become practice~\cite{lat02:reyes,lat02:lubicz} 
to eliminate the first term by forming
\bes
  B(p^2)&=&    {m^2_\pi(m_1) R(m_1) - m^2_\pi(m_2) R(m_2)
         \over m^2_\pi(m_1) - m^2_\pi(m_2)} \nonumber \\
         && + \rmO(m_\pi^2)\,, \label{e_polesubtr}
\ees
and then to determine $\zp^{-1}$ from the chiral limit of \eq{e_polesubtr}.
From a theoretical point of view, \eq{e_Rm} means that the 
renormalization constant $\zp$ (and others) 
do not exist in the massless MOM-scheme for $p=p'$.
As alternatives, one could formalize the subtraction \eq{e_polesubtr}
and include it into the {\em definition} of the renormalization scheme 
or one could choose a kinematics which truly corresponds to short distances
in position space, i.e. $p^2,(p')^2,(p-p')^2 \gg \Lambda^2_{\rm QCD}$
or one could avoid operators which couple to the pion and relate their
renormalization  to others
by Ward identities\cite{zp:givl}. The second alternative 
would certainly be a theoretically sound solution.

Concerning the numerical results alluded to above, the fact that one is dealing
with the subtraction of singular terms makes the chiral extrapolation
of the remainder numerically more difficult. In fact,
double poles have to be subtracted from 
the 4-fermion vertex functions~\cite{lat02:reyes,Bpar:SPQR1}.


\subsection{Progress using finite volume schemes \label{s_finite_vol}}
Finite volume schemes allow 
to compute the scale-dependence of $\Phiinter(\mu)$ explicitly
up to large $\mu=\rmO(100\,\GeV)$. One can then verify the onset of 
perturbative running and use the perturbative $\beta$ and $\gamma$
functions (in \eq{e_RGI2}) only a safe distance beyond that point.
In this way one computes ${\PhiRGI / \Phiinter(\mu_{\rm m})}$, \eq{e_factors},
non-perturbatively.
Apart from the early studies of renormalized 
couplings~\cite{alpha:sigma,alpha:SU2comb},
only schemes based on the \SF \cite{SF:comb} 
have been used so far. The reasons are probably that in this framework
\bsi
 \its gauge invariance is explicit,
 \its on-shell improvement is sufficient,
 \its observables show good signal/noise ratios, in\\
\hspace*{0.3cm}particular the important running coupling,
 \its one can use a massless scheme, without\\
\hspace*{0.3cm} extrapolations,
 \its perturbation theory is relatively easy and 
 \its $a$-effects are typically quite small.
\esi
Still, for new applications one should keep in mind that 
other finite volume schemes may be useful.

The central objects, which describe the $\mu$-dependence and
are computable by MC-methods are the step scaling functions.
They give the renormalized quantities $\PhiR$ at scale 
$\mu=1/(s\times L)$ as a function of those at scale $\mu=1/L$.
(In this section we use just ``R'' to denote 
the intermediate scheme). Picking a complete set of 
observables, $\PhiR_{,i}$, the
step scaling functions $F$ are
\bes
  \label{e_SSF}
  \PhiR_{,i}(\mu/s) = F_i\left(\{ \PhiR_{,j}(\mu)\}\right)\,,\;i,j=0,\ldots,M\,.
\ees 
A special r\^ole is played by the renormalized coupling
\bes
\PhiR_{,0} \equiv \gbar^2(L)\,, 
\ees
which just like in \eq{e_RG} is taken to parameterize the 
scale $\mu=1/L$. In fact, in a massless scheme, its
step scaling function
\newcommand{\uu}{{\cmag u}}
\bes
F_0 \equiv  \sigma(u) = \left.\gbar^2(sL)\right|_{u=\gbar^2(L)}
\ees
needs no further argument. 
Another example of a step scaling function is the 
case of composite operators
which mix under renormalization, where \eq{e_SSF} is realized as 
\bes
  \PhiR_{,i}({\cred\mu/s})\cbla = \sum_j {\cred\sigma_{ij}(\uu)} \PhiR_{,j}({\cred\mu})\,,
  u=\gbar^2(L)
\ees 
($\equiv F_i\left(\{ \PhiR_{,j}(\mu)\}\right)$). For an example of purely additive 
renormalization see \sect{s_HQET}, \eq{e_sigmam}.

Lattice approximants of the step scaling functions are defined at finite resolution
$a/L$ via
\bes
  \Sigma(\uu,a/L) &=& \gbar^2(sL) \nonumber\\[-1.5ex]
                  && \qquad \qquad \qquad\text{at }\gbar^2(L)=\uu \nonumber\\[-0.5ex]
  \Sigma_{ij}(\uu,a/L) &=& \sum_k Z_{ik}(sL)Z^{-1}_{kj}(L) \,. \label{e_ssf}
\ees
As indicated in the illustration for $s=2$, the r.h.s. of \eq{e_ssf}
involves quantities on lattices $L/a$ and $sL/a$ but at 
{\em the same bare parameters $g_0,m_0$} which are fixed by 
$\gbar^2(L)=\uu$ and the vanishing of the PCAC-mass~\cite{mbar:pap1}.
Thus the lattice spacing $a$ is kept fixed (and the quark mass
zero) while the 
lattice size $L$ changes by a factor $s$. In the next step
one changes $L/a$ at constant $\gbar^2$ (downwards arrows in the
graph), thus keeping $L$ fixed instead.  This process is iterated
to connect $L=L_0$ to $L=L_n=s^nL_0$. 
To approach the continuum limit, all step
scaling functions are computed for several values of $a/L$ at
fixed $u$ and extrapolated to the continuum: 
$\sigma(\uu)=\lim_{a/L\to0}\Sigma(\uu,a/L)$.
\vspace{0.3cm}

\newcommand{\bla}{\cbla}
\newcommand{\red}{\cred}
\newcommand{\gre}{\cgre}
\newcommand{\blu}{\cblu}
\newcommand{\smalllat}[1] 
{
\unitlength #1
\linethickness{0.2mm}
\bla\multiput(0,0)(0,2.0){3}{\line( 1, 0){6}}
\bla\multiput(0,0)(2.0,0){3}{\line( 0, 1){6}}
\linethickness{0.4mm}
\multiput(0,2)(2.0,0){4}{\bla\circle*{0.2}}
\multiput(0,4)(2.0,0){4}{\bla\circle*{0.2}}
\multiput(0,6)(2.0,0){4}{\red\circle*{0.2}}
\multiput(0,0)(2.0,0){4}{\red\circle*{0.2}}
\put(0,0){\blu\line(0,1){6}}
\put(6,0){\blu\line(0,1){6}}
}
\newcommand{\biglat}[1] 
{
\unitlength #1
\linethickness{0.2mm}
\bla\multiput(0,0)(0,2.0){6}{\line( 1, 0){12}}
\bla\multiput(0,0)(2.0,0){6}{\line( 0, 1){12}}
\linethickness{0.4mm}
\multiput(0,2)(2.0,0){7}{\bla\circle*{0.2}}
\multiput(0,4)(2.0,0){7}{\bla\circle*{0.2}}
\multiput(0,6)(2.0,0){7}{\bla\circle*{0.2}}
\multiput(0,8)(2.0,0){7}{\bla\circle*{0.2}}
\multiput(0,10)(2.0,0){7}{\bla\circle*{0.2}}
\multiput(0,12)(2.0,0){7}{\red\circle*{0.2}}
\multiput(0,0)(2.0,0){7}{\red\circle*{0.2}}
\put(0,0){\blu\line(0,1){12}}
\put(12,0){\blu\line(0,1){12}}
}

\unitlength 0.2cm
\begin{picture}(18,18)(1,1)

\put(1,16){\smalllat{0.035cm}}
  \linethickness{0.4mm}
  \mgt\put( 3,17){\vector(1,0){1.5}}
  \gre\put( 5,15.5){\vector(-2,-3){1.2}}
  \red\put(0.0,18.0){\small $\gbar^2(L_0)$}
\put(1,11){\smalllat{0.07cm}}
  \linethickness{0.4mm}
  \mgt\put(4.5,13){\vector(1,0){1.5}}
  \gre\put(7.5,10.5){\vector(-2,-3){2.5}}
  \red\put(15,10.5){$\small \gbar^2({\mgt 8}L_0)$}
\put(1,1){\smalllat{0.14cm}}
  \linethickness{0.4mm}
  \mgt\put( 7,5){\vector(1,0){1.5}}
\put(6.0,16){\biglat{0.035cm}}
\put(7,11){\biglat{0.07cm}}
\put(10,1){\biglat{0.14cm}}
\end{picture}


\vspace{0.3cm}

Inverting the continuum step scaling functions 
allows to climb up in energy, $\mu\to s\mu \to s^2 \mu \ldots$.
\begin{figure}[ht]
    \vspace{0 cm}
    \hspace{0 cm}
    \includegraphics[width=6.5cm]{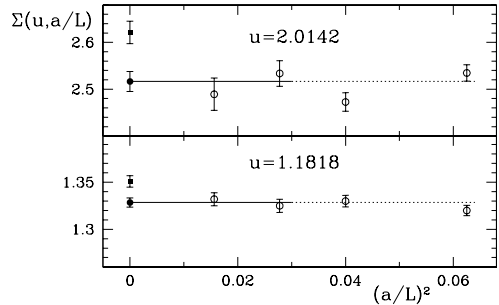}
    \vspace{-1cm }
\caption{\label{f_Sigma}\small 
Step scaling function for $\nf=2$ (circles) compared to the continuum $\sigma$
for $\nf=0$ (squares).
}
\vspace{-0.5cm}
\end{figure}
Let us illustrate recent progress in implementations
of the approach.\\
{\bf 1.} The basis for all subsequent applications is the 
running coupling. Following very closely \cite{alpha:SU3,mbar:pap1},
it has been studied also for $\nf=2$.  In addition to
last year's results \cite{alpha:letter} also $\Sigma(\uu,a/L=1/8)$
is now available~\cite{lat02:michele}. The lattice spacing dependence 
is very weak and the small effect of dynamical
fermions on the (SF-scheme) $\beta$-function is seen (\fig{f_Sigma}).
Although some of the simulations were difficult and
a modification of the importance sampling was needed to
become convinced that the configuration space is 
sampled properly in all cases, one finally has
a precision result for massless dynamical fermions!
Compared to \cite{alpha:letter} the systematic error
due to the $a/L\to0$ extrapolation is reduced, leading to~\cite{lat02:michele}
\bes
 \Lmax \Lambda_\msbar = 0.68(7)\,, \text{ where }\gbar^2(\Lmax)=5\,.
\ees  
{\bf 2.} Defining the quark mass through the renormalized PCAC relation,
its scale dependence is given by that of the renormalization
constant, $\zp$, of the 
flavor non-singlet pseudo-scalar density. In close
analogy to \cite{mbar:pap1}, its step scaling function, $\Sigmap$,
has now been computed also for $\nf=2$ and $L/a=6,8$ \cite{lat02:francesco}. 
A continuum limit requires at least one larger value of $L/a$.

%
\begin{figure}[hb]
  \begin{center}
    \vspace{0 cm}
    \hspace{0 cm}
    \includegraphics[width=6.5cm]{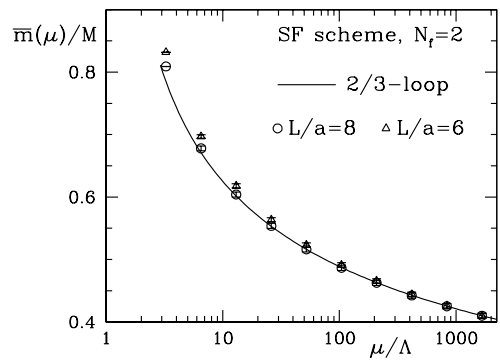}
    \vspace{-1.0 cm}
  \end{center}
\caption{\label{f_mbar}\small 
Running of quark masses for $\nf=2$. For details see \cite{lat02:francesco}.
}
\end{figure}
As a preliminary step, the running quark mass in units of the RGI-mass $M$
was evaluated from $\Sigmap(u,1/6)$ and $\Sigmap(u,1/8)$
separately.
The encouraging result is shown in \fig{f_mbar}.
Although the overall cost of the simulations
grows very rapidly with $L/a$~\cite{lat02:francesco}, using 
Hasenbusch's variant of HMC~\cite{algo:GHMC} leads to an effort of
only about a day on an APEmille crate (peak speed of 64 Gflop/s) for 
1\% precision in $\zp$ 
and for $L/a=16$. 
Thus, the step $L/a=12 \to 2L/a=24$ 
is not expected to be a problem and the continuum limit
is within reach.

Assuming $\Lambda_\msbar=\rmO(250\,\MeV)$ one finds
that the maximum box size, $\Lmax$, is large enough 
to make contact with the infinite volume physics and extract
$\Lambda_\msbar/\Fpi$ and $\Mstrange/\Fpi$. Still, performing a
continuum limit in this step may be a challenge for the future. \\
{\bf 3.} There are first results for the renormalization of parity-odd 
4-fermion operators in the SF-scheme
(quenched approximation)~\cite{lat02:carlos}. These 
are of considerable interest since their matrix elements 
give predictions for
\bsi
  \its 
       parity-odd decays when their hadronic matrix elements 
       are evaluated with the 
	standard (improved) Wilson action
  \its parity-even decays when the standard mass term is 
	replaced by the ``twisted'' one~\cite{tmqcd:comb},
       see also \cite{lat02:robertof}. We refer to this as tmQCD.
\esi
The second option is particularly interesting since
in this way $\Bk$ can be
computed without any  operator mixing.\footnote{With the Wilson action
and standard mass term, the relevant 4-fermion operator mixes
with operators of wrong chirality. This has been a significant
source of uncertainty~\cite{lat00:laurent}.}

For the definition of the operators and their renormalization
conditions, we refer to \cite{lat02:carlos}.  
A complete basis has been considered, but so far the detailed analysis
has only been done for the operator whose matrix element 
in tmQCD yields $\Bk$. 

Although in the range $a/L=1/6\, -\, 1/16$ 
and for 13 different values of $u$ a lattice 
spacing dependence of at most 5\% has been observed in the step scaling 
functions, a continuum extrapolation has to be done with care;
the operator is not $\Oa$-improved and linear and quadratic
$a$-effects may compete (see also the case of
structure functions \cite{sf:univ}). Fortunately there are a number 
of handles to control the continuum limit. 
{\bf a)} Five different intermediate schemes have been 
implemented, which all should yield the same final RGI 
matrix element  $\PhiRGI$. 
{\bf b)} For the same scheme, different regularizations
(e.g. Wilson and improved) have to extrapolate
to the same continuum step scaling function~\cite{sf:univ}.
These constraints should be sufficient to obtain accurate 
final numbers.\\
{\bf 4.} The renormalization of the static-light axial current
in the quenched approximation has been finished,
including nice continuum extrapolations of
the step scaling functions~\cite{lat02:jochen}. The overall
Z-factor, mentioned already in the introduction, has then been
used to estimate $\Fbs=261(46)\,\MeV+\rmO(1/\mbs)$ from
the bare matrix elements of \cite{stat:fnal2,stat:DMcN94}
in the unimproved theory.


\subsection{Heavy Quark Effective Theory \label{s_HQET}}

As the b-quark mass is usually beyond the cutoff $a^{-1}$
(cf. \fig{f_scales})
it is of great
interest to use the $1/m$ expansion,
implemented as an effective theory with the 
(zero-velocity) HQET Lagrangian\cite{stat:eichhill1}
($P_+\heavy=\heavy\,,\;\heavyb P_+=\heavyb\,,\;P_{\pm}= \frac{1\pm\gamma_0}{ 2}$)
 \def\vecB{{\bf B}}
 \def\vecD{{\bf D}}
\def\vecsig{{\bf \sigma}}
\def\LD{\lag{Dirac}}
\def\Dop{{\cal D}}
\newcommand{\vecg}{{\bf \gamma}}
\def\Dg{D_k\gamma_k}
\def\Lh{\lag{h}}
\def\Lhb{\lag{\bar h}}
\def\Lhhb{\lag{h \bar h}}
\bes
\lag{\rm HQET} 
    &=& \Lh^{\rm stat}  
       + \frac{1}{\cred m} \Lh^{(1)}
       +\rmO(\frac{1}{\cred m^2})\,, \\
\cmag \Lh^{\rm stat} &=&  \cmag \heavyb(D_0+m)\heavy\,. \label{e_Lstat}
\ees
The higher order terms such as 
\bes
\cbla
\Lh^{(1)} &=&  
              \heavyb(-\frac12\vecD^2 -\vecB\vecsig)\heavy\,,
\ees
are to be treated as
perturbations to the static theory defined by $\Lh^{\rm stat}$.
Then the expansion
is expected to be renormalizable order by order in $1/m$ 
(similar to chiral perturbation theory). Nevertheless, $\lag{\rm HQET}$
contains operators
of different dimension which mix under renormalization
with power divergent coefficients $\sim a^{-n}$. Estimating them 
at a given order $k$ in the coupling constant expansion, one
is left with error terms of the form
$$
a^{-n} g_0^{2k+2}\;\; {\raisebox{-.6ex}{$\stackrel{\small a\to0}{\textstyle{\longrightarrow}}$}}\;\;\infty
$$
and the continuum limit does not exist.

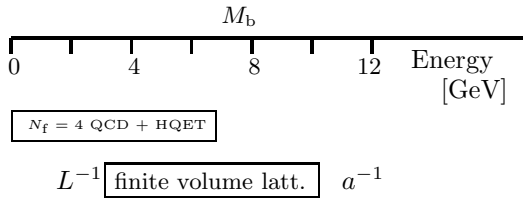
\begin{figure}[h]
\vspace{-2.3cm}
\begin{picture}(8,15)(0,0)
  \unitlength 0.4cm
  \put(0,2){\linenergy{0.4cm}}
  \put(8,2.5){\small $\Mb$}
  \put(1,-1){\ftext{\tiny \hspace{-0.1cm} $\nf=4$ QCD + HQET\hspace{-0.1cm}}}
  \put(2.5,-3){\cred $L^{-1}$}
  \put(4.1,-3){\ftext{\hspace{-0.1cm}\small finite volume latt.\hspace{-0.1cm}}}
  \put(12,-3){\cred $a^{-1}$} 
\end{picture}
\vspace{0.6cm}
\caption{\label{f_match} \small
Matching HQET and QCD  for $L\approx 0.2\fm$}
\end{figure}

This is relevant already at the level of the static theory:
a linearly divergent ($n=1$) additive mass renormalization is needed.
Unlike the case of relativistic fermions, there is no (chiral)
symmetry to fix this term. 
Thus, a continuum limit of the b-quark mass in the static approximation
requires a
{\em strategy for NP renormalization including the
power divergent subtraction}.
A viable approach was introduced
at last year's conference~\cite{lat01:rainer}. The simple idea is 
again based on the use of finite volume. It is illustrated in 
\fig{f_match}, which one should compare to \fig{f_scales}: 
if we choose the box size
$L$ appropriately, we can both
\bsi
   \its treat the b-quark as a relativistic fermion, \\
        \hspace*{0.3cm} i.e. simulate QCD
	with $a \mbeauty \ll 1$  and
   \its apply  HQET quantitatively which requires \\
        \hspace*{0.3cm} $L \mbeauty \gg 1$.\footnote{ 
These conditions are satisfied for $L\approx0.2\,\fm$,
where in comparison to a large volume simulation ($L\approx2\,\fm$)
an order of magnitude is gained in the 
lattice spacing for the same  $L/a$.}
\esi
\vspace{-0.6cm}
Identifying some suitable
observables in the two theories allows to connect the parameters of
HQET to those of QCD. This solves
the matching problem entirely non-perturbatively (in this case
the intermediate step through some RGI's (\fig{f_scales}) is
not necessary).  

The idea can be implemented in such a way
that only quantities which are defined separately in HQET and in 
QCD and which are independent of the regularization
have to be computed. 
To understand how this works, it is best to consider 
the following example.
\subsubsection{\bf The b-quark mass in the static approximation}
The renormalization of the static theory requires 
(apart from the usual one in the light sector)
a (linearly divergent) additive mass 
renormalization. In other words, 
when we insert
$ 
  m = m_\bare(g_0)
$ 
into the Lagrangian $\Lh^{\rm stat}$, \eq{e_Lstat},
with suitably chosen $m_\bare(g_0)$, then all energies
are properly renormalized and have a continuum limit.
Due to the simple form of the action, the dependence 
of the static propagator on
$m_\bare(g_0)$  is explicit and corresponds to 
a common shift of all energies. It is therefore notationally
simpler and common practice to define all observables with
$m=0$ in \eq{e_Lstat} and keep in mind that renormalized energies
are
\bes
  \Estatr=\Estat+m_\bare\,, \text{ with }m=0.
\ees 
We now want to use the B-meson mass, $\mB$, on the left hand side and
relate $m_\bare$ to the quark mass in the relativistic theory,
whose renormalization is under control\cite{mbar:pap1} to obtain the
relation between $\mB$ and the RGI b-quark mass. 
\vspace{-1.5cm}
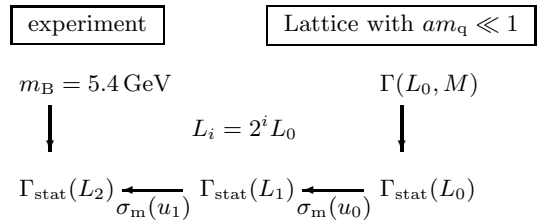
\begin{figure}[h]
\begin{picture}(8,20)(0,0)
\small
\hspace{-1.5cm}
  \unitlength 0.4cm
  \put(4,6){\ftext{experiment}}            \put(12.5,6){\ftext{Lattice with 
$a\mq\ll 1$}} 
  \put(4,4){ $\mB=5.4\,\GeV$}    \put(16,4){ $\meff(L_0,M)$} 
  \linethickness{0.3mm}\cgre\put(5.3,3.5){\vector(0,-1){1.5}}
  \linethickness{0.3mm}\cgre\put(17,3.5){\vector(0,-1){1.5}}
  \linethickness{0.3mm}\cbla
  \put(4,0.5){ $\meffstat(L_2)$}
  \put(10,0.5){ $\meffstat(L_1)$}
  \put(16,0.5){ $\meffstat(L_0)$}
  \put(15.7,0.7){\vector(-1,0){2.0}}
  \put( 9.7,0.7){\vector(-1,0){2.0}}
  \put(13.5,-0.1){$\sigmam(u_0)$}
  \put( 7.5,-0.1){$\sigmam(u_1)$}
  \put(10,2.5){\small $L_i = 2^i L_0$}
\end{picture}
\vspace{-0.5cm}
\caption{\label{f_mbstrat} \small
Connecting experimental observables to properly renormalized HQET.}
\end{figure}
\vspace{-0.3cm}

Starting from a static-light correlation function 
in the SF (see \cite{zastat:pap1} for 
the precise definition),
\bes
 \fastat(x_0) \propto \sum_{\vecy\,,\vecz}\,
  \left\langle \cblu \Astat(x)\,\zetahb(\vecy)\gamma_5\zeta_{\rm l}(\vecz)\cbla \right\rangle \\
 = \hspace*{1.5cm}\parbox{4.0cm}{

\begin{fmffile}{fastat}

\unitlength=0.3cm
\begin{fmfgraph*}(10,4)
  \fmfstraight
  \fmfleftn{i1a}{5}
  \fmfrightn{o1a}{5}
  \fmf{plain}{i1a1,i1a5}
  \fmf{plain}{o1a1,o1a5}
  \fmf{heavy,foreground=black}{i1a2,v1a1}
  \fmf{quark,foreground=black}{v1a1,i1a4}
  \fmf{phantom,tension=1.6}{v1a1,o1a3}
  \fmffreeze
  \fmfv{label.angle=0,label=$\noexpand A_0^{\rm stat}$,decor.shape=square,
        decor.filled=0,decor.size=2thick,
        foreground=black}{v1a1}
  \fmfv{label.angle=180,label=$\overline{\zeta}_{\rm h}$,
        foreground=black}{i1a2}
  \fmfv{label.angle=180,label=$\noexpand\zeta_{\rm l}$,
        foreground=black}{i1a4}
  \fmfv{label.angle=-90,label=$\noexpand x_0=0$,
        foreground=black}{i1a1}
  \fmfv{label.angle=-90,label=$\noexpand x_0=L$,
        foreground=black}{o1a1}
\end{fmfgraph*}

\end{fmffile}

}
\ees
we first define an $L$-dependent bare energy
\bes
  \meffstat(L)&=&-\frac12 (\drv0+\drvstar0)\ln[ \fastat(x_0)] \nonumber
  \\ &&\text{ at } x_0=L/2\,.  
\ees
The renormalized energy is
\bes
  \meffstatr(L) = {\meffstat}(L) + m_\bare\,. 
\ees
The same quantity computed in QCD
with finite mass of the heavy quark is denoted by
$\meff(L,M)$, where the argument $M$ is conveniently taken as the 
scheme independent
RGI mass~\cite{mbar:pap1} of the heavy quark. Matching the two theories at lowest 
order in $1/M$ is achieved by the condition
\bes
\meffstatr(L) =  
                   \meff(L,\Mb)\,. \label{e_matchstat}
\ees
To the same accuracy, 
the energy of a B-meson, 
$\Estatr$, equals the physical (experimental)
mass, $\mB$. Thus we have
\begin{figure}[htb]
  \begin{center}
    \includegraphics[width=4.5cm]{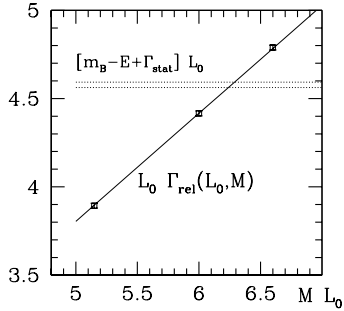}
    \vspace{-1.2cm}
  \end{center}
\caption{\label{f_gammarel} \small
Mass-dependence of $L_0 \meff(L_0,M)$, in the continuum
limit ($\nf=0$).}
\end{figure}
\vspace{-0.4cm}
\bes
 \mB &=& \Estat + m_\bare \nonumber \\
     &=& \Estat + \meff(L,\Mb) - \meffstat(L) \nonumber \\
     &=& [\Estat - \meffstat(L_n)]  \label{e_mbequ}
         \\ \nonumber
         && + 
          [\meffstat(L_n) - \meffstat(L_0)] + \meff(L_0,\Mb)\,,
\ees
for arbitrary $L_n$. Here the matching 
\eq{e_matchstat} is performed at $L=L_0$, which has to satisfy
\bes 
L_0\Mb \gg 1\,.
\ees
The two terms in ``[~~~]''-parenthesis are energy differences
in the static theory which do not
need renormalization. Their continuum limit can be taken. The 
entire quark mass
dependence is contained in the last
term $\meff(L_0,\Mb)$, defined in QCD with a relativistic 
b-quark. For a range of $M$, 
the function $\meff(L_0,M)$ can again be computed 
in the continuum limit. 
The result of a quenched study~\cite{lat01:rainer}
is shown in \fig{f_gammarel}.

Finally, 
all ingredients are present to solve \eq{e_mbequ}
for $\Mb$. The various steps of the procedure
are summarized in \fig{f_mbstrat}.

One point remains to be discussed. 
Of course the computation of
$\Estat$ 
needs a lattice large enough to accommodate a B-meson,
say of size $L \geq 1.5\,\fm$. In this step
one can then not afford lattice spacings much smaller than
$0.05\,\fm$. At the same lattice spacing 
(the same bare parameters of the theory), 
one needs to compute $\meffstat(L_n)$
to form the difference $\Estat-\meffstat(L_n)$.
Therefore also $L_n$ may not be too small.
In practice, 
$L_n=2^n L_0$, $L_0\approx0.2\,\fm$ 
with $n=2$ is sufficient\cite{lat01:rainer}.

To connect $\meffstat(L_2)$ to $\meffstat(L_0)$ it is 
furthermore convenient to use
\bes
  [\meffstat(L_2) - \meffstat(L_0)]\,L_0 =  \qquad\qquad\qquad \\
 \qquad \frac12\sigmam(u_{0}) + \frac14 \sigmam(u_{1}) 
 \quad \text{ with }
  u_i = \gbar^2(L_i), \nonumber
\ees
(the \SF coupling) and 
\bes
\sigmam(u)  &=& \lim_{a/L\to0} \Sigmam(u,a/L) \label{e_sigmam}\\
\Sigmam(u,a/L) &=& \nonumber 
                2L\,[\meffstat(2L)-\meffstat(L)]_{u=\gbar^2(L)}\,.
\ees
Unnecessary additional scales are avoided by setting the light quark mass
to zero everywhere, except for $\Estat$.
There it is set to the strange quark mass~\cite{mbar:pap3}. Correspondingly
the experimental spin averaged mass
$\mB=m_{\rm B_s}=5405\, \MeV$ enters. 

\begin{figure}[htb]
    \hspace{0 cm}
    \includegraphics[width=7.0cm]{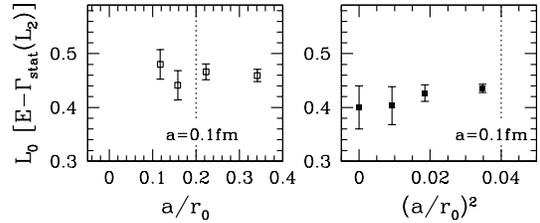}
    \vspace{-1.0 cm}
\caption{\label{f_omega_a} \small
Lattice spacing dependence of the subtracted B-meson energy
with $\Oa$ improvement (r.h.s.) and without (l.h.s.). }
\vspace{-0.3cm}
\end{figure}

The method has been tested in the quenched approximation
\cite{lat01:rainer}.  Except for 
$\Estat - \meffstat(L_2)$ all pieces entering
\eq{e_mbequ} have been extrapolated to the 
continuum. We here just mention few details
of this numerical work.
\bsi
  \its $\Oa$-improvement was employed in all steps.
	Although the improvement coefficient entering 
	the static axial current is known only perturbatively,
	this did not play a prominent r\^ole. Its effect
	on $\Sigmam$ is very small.
  \its The continuum extrapolation of $\meff(L_0,M)$ requires
	to fix $L_0$ and $z=ML_0$ while varying 
        $L_0/a$. The proper values of $g_0$ to keep $\gbar^2(L_0)=2.45$ 
	are known from \cite{alpha:SU3,mbar:pap1} and the bare subtracted 
	quark mass $\mq$ of the heavy quark is obtained from
	\bes
           z=ML_0 = L_0\, Z\,Z_M\, \mq (1+\bm a\mq) 
	\ees
	The coefficients $Z\,,Z_M\,, \bm$ are defined in \cite{mbar:pap1,impr:babp}
        and have also been computed
	non-perturbatively~\cite{impr:babp}.
	Nevertheless $\bm$ and $Z$ are needed at relatively small $g_0$,
        outside the range considered in \cite{impr:babp}. 
        They needed to be extrapolated as a function of $g_0$, 
 	causing additional uncertainties, which can be reduced
	in future work.
  \its
	Values $5.15\leq z\leq 9$ were used.  
        In this range, $\meff(L_0,M)$ turned out to be 
	an almost linear function of $z$ as naively expected.
	It could be interpolated in $z$ easily (see \fig{f_gammarel}). 
\esi
The last piece, $\Estat - \meffstat(L_2)$, has now been computed for
both the unimproved theory ($\Estat$ from \cite{stat:fnal2})
and with improvement \cite{stat:mbprelim} for various lattice spacings. Still the 
situation, 
shown in \fig{f_omega_a} is not completely satisfactory. The 
improved results show some residual dependence on the 
value of the time-separation where $\Estat$ is extracted; this leads
to large overall uncertainties.  The unimproved results
generally are somewhat higher than the improved ones,
but do have large errors for $a\leq0.1\,\fm$.
We consider 
\bes
  [\Estat-\meffstat(L_2)]\,L_0 = 0.40(4)\,,
\ees
which is shown at $a=0$, as an estimate of the continuum (quenched) result. 
Setting the scale through $\rnod$ and using
$L_0/r_0=1.436/4$~\cite{mbar:pap1}, the solution of \eq{e_mbequ} reads 
\bes 
  \Mb=6.96(8)({\cred 10})\,\GeV\,,
\ees 
where the larger part of the error originates from the uncertainties of
$Z\,,Z_M\,, \bm$.

\subsubsection{Generalizations, Perspectives \label{s_gener}}
The above example shows that the matching in finite volume
can be used in practice and for a power-divergent quantity.
We now explain briefly how it can be applied more generally.

Again consider multiplicative renormalization,
e.g. of the static-light axial current. 
In terms of 
a matrix element, $X(L)$, of the  bare operator defined 
in finite volume,
the renormalized matrix element is
\bes
  \cred \XR(L) = Z(g_0,\mu a) X(L)\,,
\ees
with a renormalization constant 
$Z(g_0,\mu a)$ defined in some independent
way (here $\mu$ and $L$ are not related).
The desired large volume renormalized matrix element is denoted by 
\bes
  \PhiR = Z(g_0,\mu a) \Phi \,.
\ees
In complete analogy to
\eq{e_mbequ} we then have\footnote{In the case of mixing, the
step scaling functions become matrices and also the first 
factor in the last line can easily be given a proper definition.}
\bes
    \PhiR = \PhistatR + \rmO(1/m) \qquad \qquad\qquad \qquad\qquad\\
        =\cbla {\PhistatR   \over \XstatR(L_2) } \;
            {\XstatR(L_2) \over  \XstatR(L_1)} \; {\XstatR(L_1) \over \XstatR(L_0)
}  \;\; 
            \XstatR(L_0) \nonumber \\
        = {\PhistatR   \over \XstatR(L_2) }\; \sigmaX(u_1)\; \sigmaX(u_0)\; \XR(L_0,M)\,.
            \nonumber \quad
\ees
The replacement of $\XstatR(L_0)$ by $\XR(L_0,M)$ is the matching in small volume.
The step scaling functions are ($Z(g_0,\mu a)$ cancels out!)
\bes
  \sigmaX(u)= \lim_{a/L\to0}\left\{\Xstat(2L) \over \Xstat(L)\right\}_{\gbar^2(L)=u}\,.
\ees
These formulae are another explicit realization of the strategy 
\fig{f_mbstrat} and are easily generalized to 
include mixing and power subtractions.

All of the above may be applied also beyond the leading order in
$1/m$. This opens the exciting perspective to obtain HQET predictions
including the first order $1/m$-correction and completely non-perturbative
in the QCD coupling. The  $1/m$-terms are then treated as insertions
in the static propagator (see e.g. \cite{stat:bbstar}). 
Of course, the matching of all quantities
has to be done consistently to order $1/m$.

A potential obstacle on this road to precision predictions for B-physics
is the generally rather bad signal-to-noise ratio of static-light 
correlation functions in large volume~\cite{stat:hashi}.
NRQCD suffers little from this problem.
Unfortunately in this theory the dimension five operator
$\heavyb(-\frac12\vecD^2)\heavy$ is part of the leading order 
Lagrangian rendering the theory unrenormalizable. It appears 
impossible to formulate NP-renormalized NRQCD in such a way 
that a continuum limit can be taken. 


\section{CONCLUSIONS, OPEN PROBLEMS \label{s_concl}}

Over the last years NP renormalization has been developed 
and applied at a steady pace. 
Major recent steps have been
\bsi
  \its the application of the MOM-scheme with fermions with exact chiral
       symmetries,  
  \its the renormalization of the static-light axial current
  \its first steps towards the renormalization of 4-fermion operators in
       the SF 
  \its the application of the SF renormalization for $\nf=2$ and
  \its the development of a strategy to renormalize HQET non-perturbatively.
\esi
The latter, proven to work in 
an example, should be developed further, in particular 
to include $1/m$-corrections. Important questions are
in how far it is necessary to implement Symanzik improvement
and how one can reduce the noise of long distance
correlation functions.
On the other hand, no particularly small 
lattice spacings are needed and thus the approach can be 
used for dynamical fermions. 
 
The computation of $\alphaSF(\mu)$ and the renormalization 
of composite operators in the SF-scheme have been shown to be 
applicable in the presence of
dynamical fermions without excessive computational requirements. Unfinished 
parts are mainly due to the connection to the 
low-energy region of the theory where
the usual problems (are the quarks light enough, the volumes large enough?)
have to be tackled.

Of course, concerning e.g. $\alphaSF(\mu)$, 
one needs $\nf=3$ dynamical fermions to make better contact with reality.
A first step in this direction has been taken by the
JLQCD/CP-PACS collaboration who have computed the important
improvement coefficient $\csw$ non-perturbatively for
$\nf=3$, using both the Wilson gauge action and the 
Iwasaki one~\cite{lat02:Aoki}.

An important open problem is whether the SF may be combined with
fermions with exact chiral 
symmetry. It should be possible to formulate
the SF with any kind of regularization, since it is (formally) defined in the 
continuum. However, no convincing formulation is known as yet. This would be 
important in order to be able to compute the last factor 
$\zinter(g_0,\mu_{\rm m} a)$, \eq{e_factors}, also in such 
regularizations without going
through yet another quantity defined with periodic boundary conditions
or in infinite volume~\cite{psibarpsi:GW1}.

{\bf Acknowledgments.} I profited from discussions with 
S. Sharpe, M. Testa, M. Papinutto and
J. Reyes on the MOM-scheme and with C. Hoelbling on overlap fermions. 
I thank  T. Bhattacharya, P. Bowman, C. Dawson, C. Hoelbling and V. Lubicz
for sending me their numbers for the quark propagator and  
U. Wolff and J. Heitger for comments on the manuscript. Last but not least 
I thank the members of the ALPHA-collaboration for enjoyable common work
and many discussions which found their way into this talk. 



\begin{thebibliography}{10}

\bibitem{lat02:francesco}
F. Knechtli et~al.,
\newblock this volume, hep-lat/0209025.
\newblock 

\bibitem{bk:JLQCD}
S. Aoki et~al.,
\newblock Phys. Rev. Lett. 81 (1998) 1778.
\newblock 

\bibitem{lat02:jochen}
J. Heitger  et~al., 
\newblock this volume, hep-lat/0209078.
\newblock 

\bibitem{stat:fnal2}
A. Duncan et~al.,
\newblock Phys. Rev. D51 (1995) 5101, hep-lat/9407025.
\newblock 

\bibitem{Lepenzie}
G.P. Lepage and P.B. Mackenzie,
\newblock Phys. Rev. D48 (1993) 2250, hep-lat/9209022.
\newblock 

\bibitem{za:comb}
M. Bochicchio et~al., 
\newblock Nucl. Phys. B262 (1985) 331;
%
M. {L\"uscher} et~al., 
\newblock Nucl. Phys. B491 (1997) 344; 
%
P. Hasenfratz et~al., 
\newblock hep-lat/0205010.
\newblock 

\bibitem{impr:lett}
K. Jansen et~al.,
\newblock Phys. Lett. B372 (1996) 275.

\bibitem{RIMOM}
G. Martinelli et~al.,
\newblock Nucl. Phys. B445 (1995) 81.
\newblock 

\bibitem{lat01:hase}
P. Hasenfratz,
\newblock Nucl. Phys. Proc. Suppl. 106 (2002) 159, hep-lat/0111023.
\newblock 

\bibitem{alpha:sigma}
M. {L\"uscher}, P. Weisz and U. Wolff,
\newblock Nucl. Phys. B359 (1991) 221.

\bibitem{mbar:pap1}
S. Capitani et~al.
\newblock Nucl. Phys. B544 (1999) 669.
\newblock 

\bibitem{reviews:renorm}
S. Sint,
\newblock Nucl. Phys. Proc. Suppl. 94 (2001) 79;
\newblock 
%
M. Testa,
\newblock Nucl. Phys. Proc. Suppl. 63 (1998) 38;
\newblock 
%
G.C. Rossi,
\newblock Nucl. Phys. Proc. Suppl. 53 (1997) 3.
\newblock 

\bibitem{lat02:reyes}
J. Reyes et~al.,
\newblock this volume, hep-lat/0209131.

\bibitem{lat02:lubicz}
V. Lubicz,
\newblock this volume.

\bibitem{lat02:laurent}
L. Lellouch,
\newblock this volume.

\bibitem{impr:comb}
K. Symanzik,
\newblock Nucl. Phys. B226 (1983) 187;
B. Sheikholeslami and R. Wohlert,
\newblock Nucl. Phys. B259 (1985) 572;
M. {L\"uscher} et~al., 
\newblock Nucl. Phys. B478 (1996) 365.

\bibitem{impr:pap3}
M. {L\"uscher} et~al.,
\newblock Nucl. Phys. B491 (1997) 323, hep-lat/9609035.

\bibitem{qprop:p1}
D. Becirevic et~al., 
\newblock Phys. Rev. D61 (2000) 114507.
\newblock 

\bibitem{lat01:bhatta}
T. Bhattacharya, R. Gupta and W.J. Lee,
\newblock Nucl. Phys. Proc. Suppl. 106 (2002) 786.
\newblock 

\bibitem{impr:offshell1}
G. Martinelli et~al.,
\newblock Nucl. Phys. B611 (2001) 311, hep-lat/0106003.
\newblock 

\bibitem{impr:offshell2}
S. Capitani et~al.,
\newblock Nucl. Phys. B593 (2001) 183, hep-lat/0007004.
\newblock 

\bibitem{lat01:sharpe}
S.R. Sharpe,
\newblock Nucl. Phys. Proc. Suppl. 106 (2002) 817, hep-lat/0110021,
\newblock 

\bibitem{lat02:giusti}
L. Giusti,
\newblock this volume.

\bibitem{GW}
P.H. Ginsparg and K.G. Wilson,
\newblock Phys. Rev. D25 (1982) 2649.
\newblock 

\bibitem{exactchi:neub}
H. Neuberger,
\newblock Phys. Lett. B417 (1998) 141, hep-lat/9707022.
\newblock 

\bibitem{exactchi:perfect}
P. Hasenfratz, V. Laliena and F. Niedermayer,
\newblock Phys. Lett. B427 (1998) 125.
\newblock 

\bibitem{exactchi:furshamir}
Y. Shamir,
\newblock Nucl. Phys. B406 (1993) 90;
\newblock 
V. Furman and Y. Shamir,
\newblock Nucl. Phys. B439 (1995) 54.
\newblock 

\bibitem{exactchi:kikunogu}
Y. Kikukawa and T. Noguchi,
\newblock hep-lat/9902022.
\newblock 

\bibitem{exactchi:martin}
M. L\"uscher,
\newblock Phys. Lett. B428 (1998) 342.
\newblock 

\bibitem{NiNi}
H.B. Nielsen and M. Ninomiya,
\newblock Phys. Lett. B105 (1981) 219.
\newblock 

\bibitem{impr:offshell3}
S. Capitani et al.,
\newblock Phys. Lett. B468 (1999) 150.
\newblock 

\bibitem{qprop:DW}
T. Blum et~al.,
\newblock Phys. Rev. D66 (2002) 014504, hep-lat/0102005.
\newblock 

\bibitem{qprop:GW}
L. Giusti, C. Hoelbling and C. Rebbi,
\newblock Phys. Rev. D64 (2001) 114508, hep-lat/0108007.
\newblock 

\bibitem{qprop:KS}
P.O. Bowman, U.M. Heller and A.G. Williams,
\newblock Phys. Rev. D66 (2002) 014505.
\newblock 

\bibitem{goldstonepole}
J.R. Cudell, A. Le~Yaouanc and C. Pittori,
\newblock Phys. Lett. B454 (1999) 105.
\newblock 

\bibitem{zp:givl}
L. Giusti and A. Vladikas,
\newblock Phys. Lett. B488 (2000) 303, hep-lat/0005026.
\newblock 

\bibitem{Bpar:SPQR1}
D. Becirevic et al., 
\newblock JHEP 04 (2002) 025.
\newblock 

\bibitem{alpha:SU2comb}
G. de~Divitiis et~al.,
\newblock Nucl. Phys. B433 (1995) 390, 
%
\newblock Nucl. Phys. B437 (1995) 447.
\newblock 

\bibitem{SF:comb}
M. {L\"uscher} et~al.,
\newblock Nucl. Phys. B384 (1992) 168; 
%
S. Sint,
\newblock Nucl. Phys. B421 (1994) 135; 
%
S. Sint,
\newblock Nucl. Phys. B451 (1995) 416.

\bibitem{alpha:SU3}
M. {L\"uscher} et~al., 
\newblock Nucl. Phys. B413 (1994) 481, hep-lat/9309005.

\bibitem{alpha:letter}
A. Bode et~al., Phys.\ Lett.\ B {\bf 515} (2001) 49.

\bibitem{lat02:michele}
M. Della~Morte et~al.,
\newblock this volume, hep-lat/0209023.
\newblock 

\bibitem{algo:GHMC}
M. Hasenbusch,
\newblock Phys. Lett. B519 (2001) 177.
\newblock 

\bibitem{lat02:carlos}
M. Guagnelli et~al., 
\newblock this volume, hep-lat/0209046.
\newblock 

\bibitem{tmqcd:comb}
R. Frezzotti et~al., 
\newblock JHEP 08 (2001) 058; 
\newblock 
%
R. Frezzotti et~al., 
\newblock JHEP 07 (2001) 048; 
\newblock 
%
M. Guagnelli et~al., 
\newblock Nucl. Phys. Proc. Suppl. 106 (2002) 320.
\newblock 

\bibitem{lat02:robertof}
R. Frezzotti,
\newblock this volume.

\bibitem{lat00:laurent}
L. Lellouch,
\newblock Nucl. Phys. Proc. Suppl. 94 (2001) 142, hep-lat/0011088.
\newblock 

\bibitem{sf:univ}
M. Guagnelli, K. Jansen and R. Petronzio,
\newblock Phys. Lett. B457 (1999) 153.
\newblock 

\bibitem{stat:DMcN94}
T. Draper and C. McNeile,
\newblock Nucl. Phys. Proc. Suppl. 34 (1994) 453, hep-lat/9401013.
\newblock 

\bibitem{stat:eichhill1}
E. Eichten and B. Hill,
\newblock Phys. Lett. B234 (1990) 511.
\newblock 

\bibitem{lat01:rainer}
J. Heitger and R. Sommer,
\newblock Nucl. Phys. Proc. Suppl. 106 (2002) 358.
\newblock 

\bibitem{zastat:pap1}
M. Kurth and R. Sommer,
\newblock Nucl. Phys. B597 (2001) 488, hep-lat/0007002.
\newblock 

\bibitem{mbar:pap3}
J. Garden et~al., 
\newblock Nucl. Phys. B571 (2000) 237.
\newblock 

\bibitem{impr:babp}
M. Guagnelli et~al.,
\newblock Nucl. Phys. B595 (2001) 44.
\newblock 

\bibitem{stat:mbprelim}
H. Molke, S.~{D\"urr}, J.~Heitger and R. Sommer,
\newblock in preparation.

\bibitem{stat:bbstar}
M. Bochicchio et~al., 
\newblock Nucl. Phys. B372 (1992) 403.
\newblock 

\bibitem{stat:hashi}
S. Hashimoto,
\newblock Phys. Rev. D50 (1994) 4639.
\newblock 

\bibitem{lat02:Aoki}
S. Aoki,
\newblock this volume.

\bibitem{psibarpsi:GW1}
P. Hernandez et~al., 
\newblock JHEP 07 (2001) 018.
\newblock 

\end{thebibliography}
\end{document}